\documentclass[aps,prl,twocolumn,superscriptaddress]{revtex4-1}
\usepackage[dvips]{graphicx}
\usepackage{amsmath,amssymb,amsthm,dsfont,url,braket,ctable,hyperref,mathtools,commath}
\usepackage[percent]{overpic}
\usepackage{booktabs}
\usepackage{multirow}
\usepackage{xcolor}  
\usepackage{colortbl}
\usepackage{tcolorbox}
\usepackage{soul}
\input{qcircuit}

\newcommand{\Tr}{\operatorname{Tr}}
\newcommand{\diag}{\operatorname{diag}}

\newcommand{\ch}[1]{\mathcal{#1}}

\newcommand{\conv}{\operatorname{conv}}

\renewcommand{\openone}{\mathds{1}}
\newcommand{\thr}[2]{W_{#1} \left( #2 \right)}
\newcommand{\hel}[2]{H_{#1} \left( #2 \right)}
\newcommand{\dst}[2]{\Delta_{#1} \left( #2 \right)}
\newcommand{\corr}[1]{\mathcal{#1}}

\newcommand{\qc}[1]{\begin{aligned} \Qcircuit  @C=6pt @R=6pt
    {#1} \end{aligned}}

\newcommand{\mc}[1]{\mathcal{#1}}
\newcommand{\bb}[1]{\mathbb{#1}}

\begin{document}

\title{Data-Driven  Inference  of  Physical  Devices:
  Theory and Implementation}

\author{Francesco Buscemi}

\email{buscemi@is.nagoya-u.ac.jp}

\affiliation{Department of  Mathematical Informatics, Nagoya
  University, Chikusa-ku, Nagoya, 464-8601, Japan}

\author{Michele \surname{Dall'Arno}}

\email{michele.dallarno@aoni.waseda.jp}

\affiliation{Faculty  of Education  and Integrated  Arts and
  Sciences,    Waseda    University,   1-6-1    Nishiwaseda,
  Shinjuku-ku, Tokyo 169-8050, Japan}

\date{\today}

\begin{abstract}
  Given  a  physical device  as  a  black  box, one  can  in
  principle  fully  reconstruct  its  input-output  transfer
  function  by  repeatedly  feeding different  input  probes
  through the  device and performing  different measurements
  on  the  corresponding  outputs.    However,  for  such  a
  complete   tomographic   reconstruction  to   work,   full
  knowledge of both input  probes and output measurements is
  required. Such  an assumption  is not  only experimentally
  demanding, but also logically questionable, as it produces
  a  circular  argument  in which  the  characterization  of
  unknown devices  appears to require other  devices to have
  been already characterized beforehand.
  
  Here, we  introduce a method to  overcome such limitations
  present in  usual tomographic  techniques.  We  show that,
  even   without  any   knowledge   about  the   tomographic
  apparatus,  it  is still  possible  to  infer the  unknown
  device to  a high degree  of precision, solely  relying on
  the  observed  data.   This  is achieved  by  employing  a
  criterion  that   singles  out  the   minimal  explanation
  compatible with  the observed data.  Our  method, that can
  be seen as a data-driven analogue of tomography, is solved
  analytically  and  implemented  as an  algorithm  for  the
  learning of qubit channels.
\end{abstract}

\maketitle

Quantum process  tomography~\cite{CN97, PCZ97,  DMP01, DL01,
  BCDFP09,  WSR19, MGSPCJRS12,  BGNMSM13, BGNRMFM17}  is the
standard  protocol   employed  to  reconstruct   an  unknown
physical device, regarded as a  black box.  In a tomographic
reconstruction, probes  are repeatedly fed as  inputs to the
black  box and  measured  at the  output.  The  input-output
transfer  function of  the  black box  can be  reconstructed
based  on  the  correlations observed  between  the  probes'
preparations  and   the  outcomes  recorded  in   the  final
measurements.  Such  a reconstruction is  reliable, however,
only  under  the  assumption  that  the  entire  tomographic
procedure, comprising the probes' preparations and the final
measurements,  is   fully  known   and  trusted.    Such  an
assumption,  beside  being  quite demanding  to  fulfill  in
practice,   is  also   unsatisfactory  from   a  fundamental
viewpoint, because  it suggests that the  knowledge required
to implement tomography can  only be obtained by recursively
resorting to another tomographic  reconstruction, and so on,
\textit{ad infinitum}.

Here,  we propose  to solve  such an  impasse by  adopting a
data-driven   (DD)   approach~\cite{Dal17,  DBBV17,   DBB17,
  DBBT18, DHBS19} to data  analysis in physical experiments.
Such an  approach relaxes any specific  assumption about the
devices involved  in the  experiment, in  the sense  that it
does  not  require  any   knowledge  of  the  input  probes'
preparations,   nor  of   the  final   measurement  settings
(measurements for short).  We then want to infer the unknown
device only on the basis of the correlations observed in the
data, without any assumption on  the apparatus that was used
to produce them,  and with respect to  any prior information
that may (or may not) be already known about the device.  We
refer to  such a task  as {\em  DD inference} of  a physical
device.

However,   the  inference   which   explains  the   observed
correlations  is, generally  speaking, not  unique: clearly,
the same  set of observed  correlations can be  explained in
many different  ways, and each possible  explanation differs
from the  others by the amount  of additional (non-observed)
correlations it  is compatible  with.  For example,  a given
set of  observed input-output  correlations could  have been
generated by a noisy channel, or by a noiseless channel with
the  same  input  and  output systems:  in  general,  it  is
impossible to  tell. This  is a typical  problem encountered
when  trying to  infer an  unknown  device on  the basis  of
partial information.  Inspired by principles such as Jaynes'
MAXENT principle~\cite{Jay57a, Jay57b}, here we also propose
to adopt a minimality criterion, according to which the best
inferential  reconstruction is  the  one  that explains  all
observed correlations and as little more as possible.

Our general ideas  can be applied, at  least numerically, to
any physical situation. However, as a concrete example, here
we analytically  solve our  method within  a large  class of
qubit channels,  which includes  many channels  of practical
interest like  all extremal qubit channels,  Pauli channels,
and  amplitude damping  channels  (this  restriction is  not
fundamental to  the algorithm  and it is  made only  for the
purpose of  obtaining analytical  results).  Even  though DD
inferential reconstruction  is insensitive to the  choice of
the computational  basis (we notice that  this limitation is
shared,    for   instance,    by   any    device-independent
protocol~\cite{Bel64,  Col06,   ABGMPS07,  GBHA10,  BCPSW14,
  PAMBGMMOHLMM10, BGLP11, HGMBAT12,  ABCB12, BBSNHMGB13}) we
show that it  is nonetheless able to provide  all except one
of  the   parameters  characterizing  the  black   box.   We
implement  our ideas  as an  algorithm for  the learning  of
qubit channels, and  test it on data generated by  the IBM Q
Experience quantum computer~\cite{IBM}.

\textit{Preliminaries}.  --- To address  the problem in full
generality,  let us  introduce  the  intuitive formalism  of
physical  circuits.  In  this framework,  time always  flows
from  left  to  right.    Single  wires  represent  physical
systems, while  double wires represent classical  inputs and
outputs that can be directly  accessed (that is, selected or
read,  respectively) by  the experimentalist.   For example,
the circuit
\begin{align*}
  \qc{i  &   &  \prepareC{\rho_i}  \cw  &   \gate{\mc{C}}  &
    \measureD{\pi_j} & \cw & j}
\end{align*}
represents the  situation in  which the  experimentalist can
choose which state $\rho_i$ is  prepared, and can read which
outcome $j$ is  output by the measurement  $\{\pi_j \}$. The
inner box labeled by $\mc{C}$ represents a channel, that is,
a physical transformation from states to states. Altogether,
the  above circuit  can be  put in  correspondence with  the
conditional probability distribution  $\{p_{j|i}\}$ it gives
rise to, in the limit of many repetitions.

For   some   given   observed   correlation   $\{p_{j|i}\}$,
conventional tomography  provides a protocol  to reconstruct
the  channel   $\mc{C}_{\textrm{T}}$  that  best   fits  the
black-box circuit
\begin{align}
  \label{eq:tomo}
  \qc{i   &   &   \prepareC{\rho_i}  \cw   &   \gate{?}    &
    \measureD{\pi_j} & \cw & j}
\end{align}
From the above, it is clear that, while the inner channel is
unknown, the probing preparation  $\{\rho_i\}$ and the final
measurement $\{\pi_j\}$ are completely known: in particular,
they must satisfy a condition of linear completeness usually
referred to as \textit{informational completeness}.

In order to move towards a data-driven approach, we
first need to consider a situation somewhat complementary to
that of conventional tomography. This is done by introducing
the set $\mc{S}(\mc{C})$ of correlations compatible with a
given channel $\mc{C}$, as follows~\cite{DBB17}
\begin{align}\label{eq:compatible}
  \mathcal{S}\left(       \mathcal{C}       \right)       :=
  \left\{\{p_{j|i}\}\equiv \quad \qc{i & & \prepareC{\ast} \cw   & \gate{\mc{C}} &
    \measureD{\ast} & \cw & j} \; \right\}\;,
\end{align}
where  each  distribution  $\{p_{j|i}  \}$  in  the  set  is
obtained by varying the  input preparation $\{\rho_i \}$ and
the  final   measurement  $\{\pi_j  \}$  (which   are  hence
represented  by the  wildcard ``$\ast$'').   The ability  to
characterize  $\mc{S}(\mc{C})$  with  respect to  any  given
prior information, that  is for all channels in  a given set
$\mc{D}$ of possible channels, is the necessary prerequisite
to perform DD inference within set $\mc{D}$.

Before turning  our attention to data-driven  inference, let
us        remark       that        Equations~\eqref{eq:tomo}
and~\eqref{eq:compatible} suggest a very simple criterion to
\textit{corroborate}~\cite{Pop59},  in  a fully  data-driven
fashion,  the reconstruction  obtained through  conventional
tomography:
\begin{tcolorbox}[title={DD Corroboration of Tomography}]
  \begin{description}
    \item[Data  Collection] Perform  conventional tomography
      as    per    Eq.~\eqref{eq:tomo}   and    denote    by
      $\mc{C}_\textrm{T}$ the reconstructed channel.
    \item[DD  Corroboration] Check  if the  distribution $\{
      p_{j|i}  \}$,  used   to  obtain  $\mc{C}_\textrm{T}$,
      belongs to $\mc{S}(\mc{C}_\textrm{T})$  or not.  If it
      does,   then  the   reconstruction  is   said  to   be
      DD-corroborated.
  \end{description}
\end{tcolorbox}
Notice  that the  above  criterion, however  obvious it  may
seem, is often not  satisfied by conventional tomography, in
which   techniques   to   cancel  errors   may   drive   the
reconstruction  away  from  observed  data.   Techniques  to
derive   \textit{self-consistent}    reconstructions,   i.e.
reconstructions always  consistent with the data,  have been
derived in  the context  of self-consistent  quantum process
tomography    and   gate-set    tomography~\cite{MGSPCJRS12,
  BGNMSM13, BGNRMFM17}, although the reconstructions therein
obtained are  not all  necessarily physical. To  address the
problem of  the non-uniqueness  of such  reconstructions, it
was therein proposed to make use of the knowledge of a given
target  $\mathcal{C}_T$, possibly  coming from  conventional
process  tomography.   Data-driven inference  represents  an
alternative approach to  derive a consistent reconstruction,
which is physical, without  the requirement of the knowledge
of such a target.

\textit{Data--driven inference.}   --- From this  moment on,
without   assuming    the   knowledge    of   reconstruction
$\mathcal{C}_T$, our aim will be to make a data--driven (DD)
inference  $\mathcal{C}_{DD}$  of  the black  box  from  the
observed correlations only.  In other words, the input probe
preparation  and   the  final  measurement   are  themselves
regarded   as  unknown   black   boxes.    In  the   circuit
representation,
\begin{align}
  \label{eq:ditomo1}
  \qc{i &  & \prepareC{?} \cw  & \gate{?} &  \measureD{?}  &
    \cw & j} \;.
\end{align}
The   crucial   idea   in    DD-inference   is   to   extend
DD-corroboration by requiring that a ``good'' reconstruction
should    be    \textit{simultaneously}   corroborated    in
\textit{any} test  that one  may perform  on the  same given
black  box.    More  precisely,  DD-inference   consists  of
obtaining   a   reconstruction   $\mc{C}_\textrm{DD}$   that
satisfies  the corroboration  criterion  for  any choice  of
$\{\rho_i\}$ and $\{\pi_j\}$. This can be done by collecting
not one,  but many  correlations $\{p_{j|i}^{(k)}  \}$. Each
such a correlation is obtained by feeding a family of states
$\{  \rho_i^{(k)} \}_i$  into the  black box  and performing
measurement $\{ \pi_j^{(k)} \}_j$ on the output states.  One
then chooses a reconstruction $\mc{C}_\textrm{DD}$ such that
$\{p_{j|i}^{(k)}  \}\in\mc{S}(\mc{C}_\textrm{DD})$, for  all
$k$.

Of course, in general such a choice is not unique.  However,
from  a physical  viewpoint,  not all  possible choices  are
equally  plausible.  Here,  we  introduce  a criterion  that
singles   out  the   ``minimal''   reconstruction  that   is
compatible with the  observed correlations $\{ p_{j|i}^{(k)}
\}$.  Let  $\mathsf{Vol}(\mc{S}(\mc{C}))$ denote  the volume
(according to  some metric)  of the set  $\mc{S}(\mc{C})$ of
correlations  compatible  with  the channel  $\mc{C}$.   Our
criterion stipulates  that the  {\em minimal  DD inferential
  reconstruction} $\mc{C}_\textrm{DD}$, with  respect to the
a  priori  information  represented  by a  set  $\mc{D}$  of
possible channels, is the solution of
\begin{align}
  \label{eq:minimal}
  \mc{C}_\textrm{DD}   :=   \underset{\substack{\mc{C}   \in
      \mc{D} \\ \{ p_{j|i}^{(k)}  \} \in \mc{S}\left( \mc{C}
      \right)}}{\arg\min}  \mathsf{Vol}\left( \mc{S}  \left(
  \mc{C} \right) \right)\;.
\end{align}
Equation~\eqref{eq:minimal} provides, in  a data-driven way,
the reconstruction $\mc{C}_{\textrm{DD}}$ which explains all
the observed correlations $\{ p^{(k)} \}$ and as little more
as  possible.

The protocol for DD inference is summarized in the following
box:
\begin{tcolorbox}[title={DD Inference}]
  \begin{description}
  \item[Data Collection] Choose families $\{ \rho_i^{(k)}\}$
    of states and measurements $\{ \pi_j^{(k)} \}$ (ideally,
    sample  uniformly over  state  and measurement  spaces).
    For any $k$, do the following:
    \begin{enumerate}
    \item Feed $\{\rho_i^{(k)}\}$ into black box;
    \item  Measure $\{\pi_j^{(k)}\}$  on the  output of  the
      black box;
    \item Collect correlation $p := \{ p_{j|i}^{(k)} \}$;
    \end{enumerate}
  \item[DD Inference]  Solve Eq.~\eqref{eq:minimal}  (in the
    quantum    case,   using    the   characterization    of
    $\mc{S}(\mc{C})$  provided in  the  next section),  thus
    obtaining    the     DD    inferential    reconstruction
    $\mc{C}_{\textrm{DD}}$.
  \end{description}
\end{tcolorbox}

While in  the data  collection stage the  experimentalist is
assumed to have  full knowledge of the  apparatus (i.e., the
states to prepare and the  measurements to perform are known
and  trusted by  the experimentalist),  the inference  stage
does not  require such a knowledge  at all, as it  uses only
the correlations obtained without  any reference about which
states and measurements produced  such correlations. In this
sense,   the  narrative   can  be   given  as   if  a   good
experimentalist, perfectly knowing her laboratory, is trying
to convince a very stubborn theoretician, who does not trust
anything apart from the bare data, about the availability of
a particular channel in her laboratory.

{\em The  quantum case}.  ---  As discussed in  the previous
secion, the ability to characterize the set $\mc{S}(\mc{C})$
with respect to any given prior information, that is for all
channels $\mc{C}$ in a given  set $\mc{D}$, is the necessary
prerequisite  to perform  DD  inference. Of  course, such  a
characterization can be obtained,  at least numerically, for
any set $\mc{D}$. The main result of this section (proved in
the Supplemental Material) is  to analytically obtain such a
characterization for a relevant class of quantum channels.

Within  quantum theory,  any  state  $\rho$ and  measurement
$\{\pi_j\}$ are represented  by a density matrix,  that is a
positive semi-definite  operator, and by  a POVM, that  is a
family  of positive-semidefinite  effects such  that $\sum_j
\pi_j =  \openone$, respectively.   Any channel  $\mc{C}$ is
represented by a completely-positive trace-preserving linear
map.   In   the  tomographic   setup,  the   probability  of
measurement outcome $j$ given input $i$ is given by the Born
rule, that is
\begin{align*}
  p_{j|i} = \Tr\left[ \mc{C}(\rho_i) \pi_j \right].
\end{align*}

\begin{figure}[h!]
  \begin{overpic}[width=.75\columnwidth]{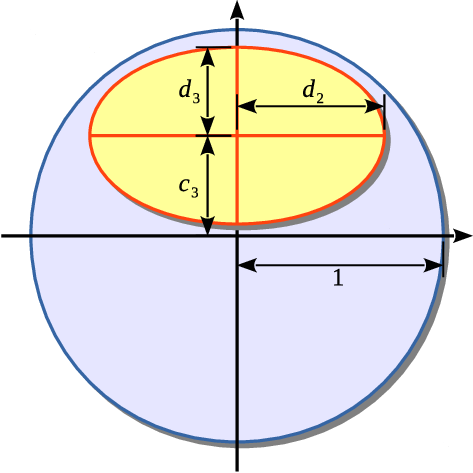}
  \end{overpic}
  \caption{{\bf   Parametrization  of   $\bb{D}_2$-Covariant
      Channels}   Geometrically,    it   turns    out   that
    $\bb{D}_2$-covariant  channels are  those  that map  the
    Bloch-sphere (the set of qubit states) into an ellipsoid
    traslated along  one of  its own axis.   Hence, up  to a
    choice  of  the computational  basis  in  the input  and
    output  spaces  (technically, up  to  (anti)-unitaries),
    $\bb{D}_2$-covariant  channels are  parametrized by  the
    lengths of  the three semi-axis $d_1$,  $d_2$, and $d_3$
    of such  an ellipsoid,  and by the  length $c_3$  of the
    traslation vector.}
  \label{fig:parametrization}
\end{figure}

The class  of quantum channels we  focus on here is  that of
qubit dihedrally-covariant  ($\bb{D}_2$-covariant for short)
channels (see Fig.~\ref{fig:parametrization} for a pictorial
representation of their action on  the state space).  Such a
class is  particularly relevant for applications,  since any
extremal  qubit  channel   is  $\bb{D}_2$-covariant,  as  it
immediately  follows from  Refs.~\cite{KR01, RSW02}.   Also,
this class includes any Pauli and amplitude-damping channel.

We     adopt     the    following     parametrization     of
$\bb{D}_2$-covariant channels.  For any $\bb{D}_2$-covariant
channel  $\mc{C}$,  let  $A_{j,i}   :=  \frac12  \Tr  \left[
  \sigma_j \mc{C}\left(  \sigma_i \right) \right]$  and $b_j
:=  \frac12  \Tr  \left[  \sigma_j  \mc{C}  \left(  \openone
  \right)\right]$.    For   $V^T   A   U$   singular   value
decomposition of $A$,  let $ \vec{d} := \diag(V^T  A U)$ and
$\vec{c} = V^T \vec{b}$.  We denote:
\begin{itemize}
\item the  only non-null entry of  $\vec{c}$ with $c_3$,
\item the  corresponding entry of $\vec{d}$  with $d_3$,
\item  the remaining  entries  of $\vec{d}$  with $d_2$  and
  $d_1$, so that $d_2 \ge d_1$.
\end{itemize}
In   other   words,   there   exists  a   basis   in   which
$\bb{D}_2$-covariant  channel  $\mathcal{C}$   acts  as  the
following linear transformation:
\begin{align*}
  \mathcal{C} \equiv
  \begin{bmatrix}
    1 & 0 & 0 & 0\\
    0 & d_1 & 0 & 0\\
    0 & 0 & d_2 & 0\\
    c_3 & 0 & 0 & d_3
  \end{bmatrix}.
\end{align*}

Such   a  parametrization   has  an   intuitive  geometrical
interpretation, as depicted in Fig~\ref{fig:parametrization}
(for further details see the Supplemental Material).

Let  us   provide  a   parametrization  for  the   space  of
correlations   (further  details   can  be   found  in   the
Supplemental  Material).  Consider  the following  matrices,
which   are   pairwise    orthonormal   according   to   the
Hilbert--Schmidt product:
\begin{align*}
  U = \frac12 \begin{bmatrix} 1 &  1 \\ 1 & 1 \end{bmatrix},
  \quad  X  =  \frac12  \begin{bmatrix}   1  &  -1  \\  1  &
    -1 \end{bmatrix}, \quad Y  = \frac12 \begin{bmatrix} 1 &
    -1 \\ -1 & 1 \end{bmatrix}.
\end{align*}
For $|x+y| \le  1$ and $|x-y| \le 1$,  we parametrize binary
conditional probability distributions with coordinates $(x, y)$ as follows
\begin{align*}
  p = \begin{bmatrix}
    p_{1|1} & p_{2|1}\\ p_{1|2} & p_{2|2}
  \end{bmatrix} =
  U + x X  + y Y.
\end{align*}
For given  correlation $p$,  parameters $x$  and $y$  can be
easily found, as follows
\begin{align}
  \label{eq:axis}
  x =  \Tr\left[ X^T p  \right], \quad  y = \Tr\left[  Y^T p
    \right].
\end{align}

It  turns out  (see the  Supplemental Material  for details)
that  the set  $\mc{S}(\mc{C})$  of correlations  compatible
with any given $\bb{D}_2$-covariant channel $\mc{C}$ is then
given by
\begin{align*}
  \mc{S}(\mc{C})  =  \conv\left[   (\pm1,  0),  \;  \corr{E}
    \right],
\end{align*}
where $\conv$  denotes the convex hull  and $\mc{E}$ denotes
the intersection of an ellipse  with the stripe $|x| \le c_3$
given by
\begin{align*}
  \mc{E} := \left\{ (x,y) \; \Big|  \; (x, y) Q (x, y)^T \le
  1 \; \wedge \;  |x| <  c_3
  \right\},
\end{align*}
where
\begin{align*}
  Q := \begin{cases} \diag\left( 0, \frac1{d_3^2} \right), &
    \textrm{    if   }    d_2    \le   d_3,\\    \diag\left(
    \frac{d_2^2-d_3^2}{d_2^2c_3^2}, \frac1{d_2^2} \right), &
    \textrm{ if } d_2 > d_3.
  \end{cases}
\end{align*}
Hence, by explicit computation one has
\begin{align*}
  \mathsf{Vol}\left(\mathcal{S}\left(\mathcal{C}\right)\right)
  = \begin{cases} d_3, & \textrm{ if  } d_2 \le d_3,\\ d_3 +
    \frac{d_2^2    c_3}{\sqrt{d_2^2   -    d_3^2}}   \arcsin
    \frac{\sqrt{d_2^2 - d_3^2}}{d_2}, &  \textrm{ if } d_2 >
    d_3.
    \end{cases}
\end{align*}

\begin{figure}[h!]
  \begin{overpic}[width=.75\columnwidth]{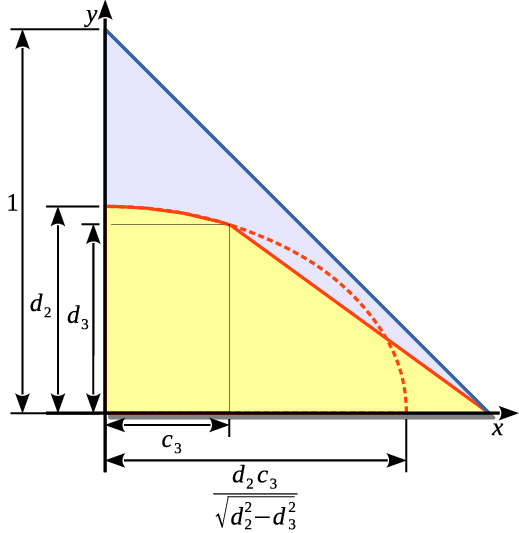}
  \end{overpic}
  \caption{A  geometrical representation  of the  mapping of
    parameters $\vec{d}$  and $c_3$, which  characterize any
    $\bb{D}_2$-covariant             channel            (see
    Fig.~\ref{fig:parametrization}),   into  the   space  of
    binary   correlations   (blue   area).    Adopting   the
    parametrization described in the  main text, points $(0,
    0)$, $(1,  0)$, and $(0,  1)$ correspond to  the uniform
    distribution  $p_{j|i} =  1/2$ for  any $i,  j$, to  the
    maximally unbalanced distribution $p_{0|i}  = 1$ for any
    $i$,   and   to   perfect  discrimination   $p_{j|i}   =
    \delta_{i,j}$, respectively. The set $\mc{S}(\mc{C})$ of
    correlations      compatible     with      any     given
    $\bb{D}_2$-covariant channel  $\mc{C}$ (yellow  area) is
    given by the intersection  of an ellipsoid (orange line)
    with  the stripe  $|x|  \le c_3$,  in  convex hull  with
    points $(\pm1, 0)$.  Since $\mc{S}(\mc{C})$ is symmetric
    under  sign flip  of coordinates  $x$ and/or  $y$ (which
    correspond to permutation of input/output indexes), only
    the  positive quadrant  is  represented.  Plot axis  are
    given by Eqs.~\eqref{eq:axis}.}
  \label{fig:characterization}
\end{figure}

This        situation        is        illustrated        in
Fig.~\ref{fig:characterization}.

As shown in the  Supplemental Material, among the parameters
$\vec{d}$   and   $c_3$    that   characterize   any   given
$\bb{D}_2$-covariant  channel $\mc{C}$,  which  ones can  be
reconstructed  by  DD  inference  depends on  the  value  of
function $\mu(\mc{C})$ given by
\begin{align*}
  \mu(\mc{C})              :=              \frac{1-c_3}{c_3}
  \frac{d_2^2-d_3^2}{d_3^2}.
\end{align*}
One has the following regimes:
\begin{description}
\item[Regime  $\mu(\mc{C}) \le  0$] reconstruction  of $c_3$
  and $d_3$;
\item[Regime $0 < \mu(\mc{C}) < 1$] reconstruction of $d_2$,
  $d_3$, $c_3$;
\item[Regime  $1 \le  \mu(\mc{C})$] reconstruction  of $d_2$
  and $\frac{d_2^2-d_3^2}{c_3^2}$.
\end{description}

{\em Learning of qubit  channels}.  ---As an application, we
implement  our ideas  as an  algorithm for  the learning  of
qubit  channels, and  we  test them  on data  experimentally
generated  by the  IBM Q  Experience quantum  computer.  Our
experiment  is  programmed  in  the  Open  Quantum  Assembly
language~\cite{CBSB17}  and  run  on  the  IBM  QX4  quantum
chip~\cite{IBM}.   For a  target qubit  channel, we  perform
conventional  (implementation-dependent) process  tomography
and  its  data-driven  counterpart,   as  discussed  in  the
previous sections.  We show that,  in this case, the results
of  conventional tomography  and  data-driven inference  are
compatible with high accuracy.

As   a   case   study,   we    chose   from   the   set   of
$\bb{D}_2$-covariant   channels    the   amplitude   damping
~\cite{NC00}  channel  $\mc{A}_{1/2}$ with  noise  parameter
$1/2$.  According to the  notation developed in the previous
section, such  a channel is  uniquely identified, up  to the
choice of the computational  basis, by parameters $\vec{d} =
(1/\sqrt{2},  1/\sqrt{2}, 1/2)$  and  $c_3 =  1/2$.  Due  to
noise, the  actual implementation $\mc{C}$ will  turn out to
be  quite  far  from the  ideal  prediction  $\mc{A}_{1/2}$.
However, this is no concern in this context since our aim is
to   compare   data-driven   inference   with   conventional
tomography,  rather  than  with the  ideal  prediction.   An
implementation  -- that  is,  a Stinespring  dilation --  of
$\mc{A}_{1/2}$  in  terms  of single-  and  two-qubit  gates
directly  supported by  the IBM  back  end is  given in  the
dashed box below,  where we also show probes  $\{ \rho_i \}$
and measurement $\{\pi_j\}$:
\begin{align*}
  \qc{\lstick{i} & \prepareC{\rho_i} \cw  & \qw & \gate{H} &
    \qw & \gate{\sigma_X} & \gate{H} & \gate{\sigma_X} & \qw
    & \qw & \qw & \measureD{\pi_j} & \cw & \rstick{j} \\ &&&
    \prepareC{\ket{0}}&  \gate{R_Y^\dagger} &  \control \qwx
    \qw  &   \gate{R_Y}  &  \control   \qwx  \qw  &   \qw  &
    \measureD{\openone} \gategroup{1}{4}{2}{10}{3pt}{--}}
\end{align*}
Here, $R_Y  := \exp(-i\pi \sigma_Y/8)$, $H$,  and $\sigma_X$
represent a  $\pi/4$-rotation around $Y$-axis,  the Hadamard
gate, and the NOT gate, respectively.

As probes $\{  \rho_i \}$ and measurement $\{  \pi_j \}$, we
chose the eigenstates of the Pauli matrices $\vec{\sigma} :=
\{ \sigma_1  := \sigma_X,  \sigma_2 := \sigma_Y,  \sigma_3 =
\sigma_Z  \}$.  Let  us denote  with $\ket{\sigma_k^i}$  the
eigenvector of  $\sigma_k$ corresponding to  eigenvalue $+1$
($i =  0$) and $-1$  ($i = 1$).   The set of  projectors $\{
\ket{\sigma_k^i}     \!\!     \bra{\sigma_k^i}     \}$    is
informationally   complete  and   is   proportional  to   an
informationally  complete measurement,  hence is  a suitable
choice  for   a  tomographic  probe  and   measurement.   An
implementation of $\{ \ket{\sigma_k^i} \}$ in terms of gates
supported by the IBM back end is given by
\begin{align*}
  \ket{\sigma_k^i}        =       S^{\delta\left(k,2\right)}
  H^{1-\delta\left(k,3\right)}
  \sigma_X^{\delta\left(i,1\right)}\ket{0},
\end{align*}
where $S := \sqrt{\sigma_Z}$ represents the Phase gate.

Hence  we  collect a  family  $\{  p^{(k,l)} \}$  of  binary
conditional      probability       distributions,      where
$p_{j|i}^{(k,l)}       :=      \bra{\sigma_l^j}       \mc{C}
(\ket{\sigma_k^i}\!\!\bra{\sigma_k^i})\ket{\sigma_l^j}$. Each
distribution $p^{(k,l)}$  is obtained as the  frequencies of
outputs $j$ given  inputs $i$ over $8192$ runs.   We use the
same  raw  data  for  conventional  tomography  as  well  as
data-driven inference of channel $\mc{C}$.

Conventional     tomography    produces     the    following
reconstruction $\mc{C}_{\textrm{T}}$ for channel $\mc{C}$:
\begin{align}
  \label{eq:usualtomo}
  \mc{C}_{\textrm{T}} : 
  \begin{cases}
    \vec{d} = \left(0.573, 0.603, 0.430 \right),\\ \vec{c} =
    \left(0.134, 0.0674,  0.508 \right) \simeq  \left(0, 0,
    0.508 \right),
  \end{cases}
\end{align}
where  setting   to  zero   the  entries  $c_1$   and  $c_2$
corresponds  to   projecting  $\mc{C}$   into  the   set  of
$\bb{D}_2$-covariant  channels.   Such an  approximation  is
compatible  with the  nominal  errors  associated with  each
two-qubit gate and  measurement for the IBM  back end, which
are around $2\%$ and $5\%$, respectively (we recall that our
setup includes two of the former and one of the latter).

We proceed  now to  discuss data-driven  inference of
channel  $\mc{C}$. By  solving the  optimization problem  in
Eq.~\eqref{eq:minimal}   we  have   the  following   minimal
DD-inference $\mc{C}_{\textrm{DD}}$ for channel $\mc{C}$:
\begin{align}
  \label{eq:ditomo}
  \mc{C}_{\textrm{DD}} : 
  \begin{cases}
  \vec{d}   =  \left(0.313   \le  d_1   \le  0.606,   0.606,
  0.437\right),\\ \vec{c} = \left(0, 0, 0.481 \right).
  \end{cases}
\end{align}
As  discussed in  the  previous section,  data-driven
inference is unable to uniquely reconstruct parameter $d_1$.
However, the upper and lower bounds in Eq.~\eqref{eq:ditomo}
immediately   follow  from   the  requirement   of  complete
positivity for channel  $\mc{C}_{\textrm{DD}}$.  Notice that
each parameter in  Eq.~\eqref{eq:ditomo} deviates from those
in Eq.~\eqref{eq:usualtomo} by $6\%$ or less.

\begin{figure}[h!]
  \begin{overpic}[width=.75\columnwidth]{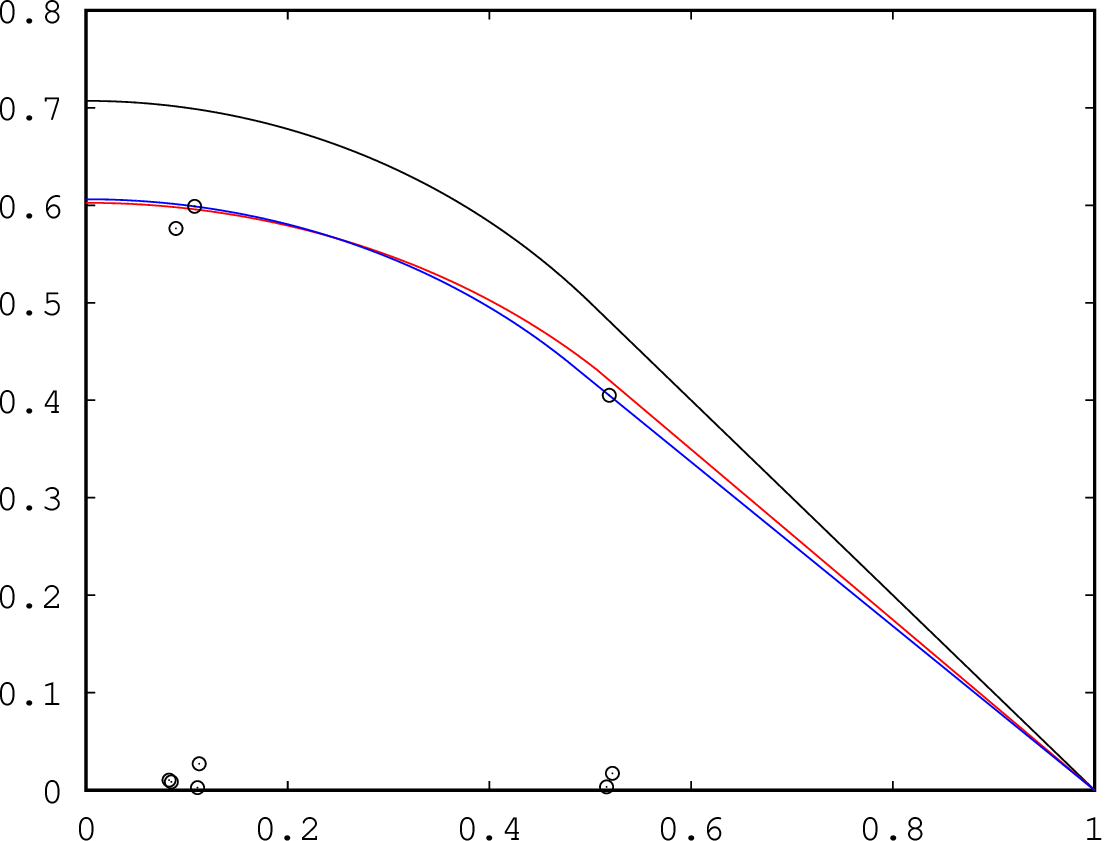}
    \put (100,  4) {$x$}
    \put (6, 78)  {$y$}
    \put   (55,    70)   {$\mathcal{S}    (
           \mathcal{A}_{1/2} )$ (black line)}
    \put (55,  65) {$\mathcal{S}  ( \mathcal{C}_{\textrm{T}}
      )$ (red line)}
    \put   (55,    60)   {$\mathcal{S}    (
           \mathcal{C}_{\textrm{DD}} )$ (blue line)}
  \end{overpic}
  \caption{Representation         of        the         sets
    $\mc{S}(\mc{C}_{\textrm{T}})$     (red     line)     and
    $\mc{S}(\mc{C}_{\textrm{DD}})$     (blue    line)     of
    correlations       compatible        with       channels
    $\mc{C}_{\textrm{T}}$     and    $\mc{C}_{\textrm{DD}}$,
    respectively, with the  parametrization discussed in the
    previous  section.   Channels $\mc{C}_{\textrm{T}}$  and
    $\mc{C}_{\textrm{DD}}$    have    been    obtained    by
    conventional tomography and  by data-driven inference of
    channel  $\mc{C}$,   respectively.  The  raw   data  $\{
    p_{j|i}^{(k,l)}  \}$ used  for both  procedures is  also
    depicted   (round    marks),   along   with    the   set
    $\mc{S}(\mc{A}_{1/2})$  of correlations  compatible with
    the  ideal  amplitude   damping  channel  (black  line).
    Plot axis are given by Eqs.~\eqref{eq:axis}.}
  \label{fig:experiment}
\end{figure}

We conclude  by comparing the  results $\mc{C}_{\textrm{T}}$
and  $\mc{C}_{\textrm{DD}}$ of  conventional tomography  and
data-driven   inference,   respectively.   The   sets
$\mc{S}(\mc{C}_{\textrm{T}})$                            and
$\mc{S}(\mc{C}_{\textrm{DD}})$  of  correlations  compatible
with each channel are depicted in Fig.~\ref{fig:experiment}.
As a  measure of distance between  $\mc{C}_{\textrm{T}}$ and
$\mc{C}_{\textrm{DD}}$ we  chose the difference  between the
Euclidean volume  $\mathsf{Vol}$ (an  area in this  case) of
the    union   and    the   intersection    of   the    sets
$\mc{S}(\mc{C}_{\textrm{T}})$                            and
$\mc{S}(\mc{C}_{\textrm{DD}})$ (usually  referred to  as the
symmetric  difference pseudometrics).  We  normalize such  a
distance by the maximum of the two volumes, thus obtaining:
\begin{align*}
  d(\mc{C}_0,                   \mc{C}_1)                  =
  \frac{\mathsf{Vol}\left(\mc{S}\left(\mc{C}_0\right)   \cup
    \mc{S}\left(\mc{C}_1\right)\right)                     -
    \mathsf{Vol}\left(\mc{S}\left(\mc{C}_0\right)       \cap
    \mc{S}\left(\mc{C}_1\right)\right)}
       {\max\left\{\mathsf{Vol}\left(\mc{S}\left(\mc{C}_0\right)\right),
         \mathsf{Vol}\left(\mc{S}\left(\mc{C}_1\right)\right)\right\}}.
\end{align*}
In    our    case    we    obtain    $d(\mc{C}_{\textrm{T}},
\mc{C}_{\textrm{DD}}) \simeq 0.0164 < 2\%$.

{\em Conclusion}. --- In this  work we addressed the problem
of reconstructing  the input-output  transfer function  of a
physical device given as a black-box.  We provided a general
protocol   for   the   data-driven  inference   of   unknown
physical-devices, based  on a minimality  principle inspired
by Jaynes' MAXENT principle. We analytically solved the case
of  dihedrally-covariant qubit  channel, which  includes any
extremal qubit channel, any Pauli channel, and any amplitude
damping  channel. Finally,  we implemented  our ideas  as an
algorithm  for the  learning of  qubit channels,  and tested
them with  data generated  by the  IBM Q  Experience quantum
computer.  The present ideas were  also recently put to test
with a quantum-optical implementation by the present authors
and others in Ref.~\cite{APCSCBDS18}.

{\em Acknowledgments}.   --- F.B. acknowledges  support from
the  Japan  Society  for  the Promotion  of  Science  (JSPS)
KAKENHI, Grant  No.  19H04066.  M.  D.  acknowledges support
from the  MEXT Quantum  Leap Flagship Program  (MEXT Q-LEAP)
Grant  No.    JPMXS0118067285.   Both   authors  acknowledge
support from the program  for FRIAS-Nagoya IAR Joint Project
Group.

{\em Author  contributions}. --- M.D.  and  F.B.  developed
the  theoretical ideas  and calculations,  and prepared  the
presentation.

{\em  Conflicts  of  interest}.  ---  The  authors  have  no
potential financial or non-financial conflicts of interest.

{\em Data Availability Statement}. --- The data that support
the  findings of  the experimental  part of  this study  are
available  from  the  corresponding author  upon  reasonable
request.

\clearpage

\section{Supplemental material}

In this section  we derive the theoretical  results on which
this work  is based.  First, we  introduce a parametrization
for binary conditional probability distributions and discuss
its     symmetries.      Then,    we     introduce     qubit
dihedrally-covariant channels and  discuss their covariances
under  unitary and  anti-unitary transformations.   Next, we
derive   the   set   of   binary   conditional   probability
distributions  which are  compatible  with  any given  qubit
dihedrally  covariant  channel.    Finally,  we  derive  the
equivalence classes  of qubit dihedrally  covariant channels
which are data-drivenly indistinguishable.

\subsection{Binary        conditional       probability
  distributions}
\label{sec:distributions}

Let  us first  introduce  a  convenient parametrization  for
binary conditional probability  distributions.  To this aim,
we introduce the following matrices:
\begin{align*}
  U := \frac12 \begin{pmatrix} 1 & 1 \\ 1 & 1 \end{pmatrix}, \quad X
  := \frac12 \begin{pmatrix} 1 & -1 \\ 1 & -1 \end{pmatrix}, \quad Y
  := \frac12 \begin{pmatrix} 1 & -1 \\ -1 & 1 \end{pmatrix},
\end{align*}
which are  orthonormal with  respect to  the Hilbert-Schmidt
product.    Then,   one    has   the   following   Cartesian
parametrization    for   binary    conditional   probability
distributions
\begin{align}
  \label{eq:cartesian}
  p_{j|i} = \begin{pmatrix}
    p_{1|1} & p_{2|1}\\
    p_{1|2} & p_{2|2}
  \end{pmatrix} =
             U + x X + y Y.
\end{align}
where  $|x+y| \le  1$ and  $|x-y| \le  1$. Of  course, given
distribution $p$, parameters $x$ and $y$ can be easily found
as follows:
\begin{align*}
  x = \Tr\left[ X^T p \right], \quad y = \Tr\left[ Y^T p \right]. 
\end{align*}

Notice  that permuting  the  inputs or  the  outputs of  $p$
correspond to the  transformations $(x, y) \to  (x, -y)$ and
$(x, y) \to  (-x, -y)$, respectively. Hence,  without loss of
generality in  the following we  take $x,  y \ge 0$,  and we
will later recover the general case by considering symmetries
around the $x$ and $y$ axis.

\subsection{Qubit dihedrally-covariant channels}
\label{sec:channels}

Let us turn  now to the parametrization  of qubit dihedrally
covariant    channels.     In   the    usual    Bloch-sphere
representation,  any qubit  state  or  unit-trace effect  is
represented as
\begin{align}
  \label{eq:bloch}
  \rho_{\vec{v}} =  \frac12 \left( \openone  + \vec{\sigma}^T
  \cdot \vec{v} \right),
\end{align}
where  $\vec\sigma  =  (\sigma_1 \equiv  \sigma_X,  \sigma_2
\equiv  \sigma_Y,  \sigma_3  \equiv \sigma_Z)$  denotes  the
vector   of  Pauli   matrices  and   $|\vec{v}|_2  \le   1$.
Accordingly, any qubit channel can be represented as
\begin{align*}
  \mathcal{C}_{A, \vec{b}} \left(  \rho_{\vec{v}} \right) :=
  \frac12 \left[  \openone +  \vec{\sigma}^T \cdot  \left( A
    \vec{v} + \vec{b} \right) \right],
\end{align*}
where $A_{i,j} = \frac{1}{2}  \Tr \left[ \sigma_i \ch{C}_{A,
    \vec{b}}  \left( \sigma_j  \right) \right]$  and $b_i  =
\frac{1}{2} \Tr  \left[ \sigma_i \ch{C}_{A,  \vec{b}} \left(
  \openone \right) \right]$.  This parametrization for qubit
channels was exploited in Refs.~\cite{KR01,RSW02}.

Let  $\mathcal{U}$  and   $\mathcal{V}$  be  two  qubit
unitary  or  anti-unitary   transformations  such  that
$\ch{V} \circ  \ch{C}_{A, \vec{b}}  \circ \ch{U}$  is a
channel.  Then by explicit computation one has
\begin{align*}
  \ch{V}   \circ   \ch{C}_{A, \vec{b}}  \circ   \ch{U}   =
  \ch{C}_{V^T A U, V^T \vec{b}},
\end{align*}
where $U,  V$ are  proper rotation matrices  if and  only if
$\mathcal{U}$ and $\mathcal{V}$ are unitary transformations,
and  improper  rotation  matrices (that  is,  rotations  and
reflections) otherwise.   By choosing  for $U$ and  $V$ some
rotation matrices  such that $D =  V^T A U$ is  diagonal, we
put $D  = \diag(d_1,  d_2, d_3)$ and  $\vec{c} =  (c_1, c_2,
c_3) := V^T \vec{b}$. Notice  that such matrices $U$ and $V$
are not unique.  By  explicit computation, the Choi operator
$R$ of $\ch{C}_{D, \vec{c}}$ is given by
\begin{align*}
  R = \begin{pmatrix}  1+c_3+d_3 & c_1 - i c_2 &  0 & d_1 +  d_2\\ c_1 + i c_2 &
    1-c_3-d_3  & d_1-d_2  & 0\\  0 &  d_1-d_2 &  1+c_3-d_3 &
    c_1-i c_2\\ d_1+d_2 & 0 & c_1 + i c_2 & 1-c_3+d_3
    \end{pmatrix}.
\end{align*}

Qubit channel $\ch{C}_{D,  \vec{c}}$ is dihedrally covariant
if and  only if two entries  of $\vec{c}$ are zero.   In the
following  we  will   consider  qubit  dihedrally  covariant
channels  only.  Notice  that  a  cyclic permutation  matrix
(that is,  a rotation  matrix) in $V$  and $U$  permutes the
entries of $D$ and $\vec{c}$. Hence, we take without loss of
generality $c_1 = c_2 = 0$.  Replacing this condition in the
Choi   operator,  the   following  condition   for  complete
positivity immediately follows
\begin{align}
	\label{eq:cp}
	\begin{cases}
    	d_3 + \sqrt{\left( d_1 - d_2 \right)^2 + c_3^2} \le 1,\\
        -d_3 + \sqrt{\left( d_1 + d_2 \right)^2 + c_3^2} \le 1.
    \end{cases}
\end{align}

Notice that  without loss of  generality we can  take $d_2$,
$d_3$,  and  $c_3$  non-negative.   This  can  be  shown  as
follows.   First, if  $c_3 <  0$, a  $\pi$-rotation in  $V$,
around  the eigenvector  corresponding  to eigenvalue  $d_1$
flips $c_3$'s sign  (it also flips $d_2$  and $d_3$'s signs,
but this  is irrelevant).  Hence without  loss of generality
$c_3  \ge 0$.   Analogously,  if $d_2  < 0$  or  $d_3 <  0$,
respectively, a $\pi$-rotation in $U$ around the eigenvector
corresponding  to eigenvalue  $d_3$ or  $d_2$, respectively,
flips $d_2$  or $d_3$'s  signs, respectively (notice  such a
rotation  does  not  flip  any sign  in  $\vec{c}$).   Hence
without loss of generality $d_2 \ge 0$ and $d_3 \ge 0$.

Notice that without  loss of generality we  can further take
$d_1$ non-negative. This  can be shown as  follows. The sign
of $d_1$ can be flipped -- without side effects on the other
parameters -- by a reflection  in $U$ around the eigenvector
corresponding to  eigenvalue $d_1$.  Here we  show that such
an  anti-unitary   transformation  preserves   the  complete
positivity.  Indeed, the l.h.s.   of the first inequality in
Eq.~\eqref{eq:cp} does not increase  if $-|d_1|$ is replaced
by $|d_1|$ (recall that $d_2  \ge 0$).  Also, the l.h.s.  of
the second  inequality in Eq.~\eqref{eq:cp} with  $|d_1|$ is
not larger  than the  l.h.s.  of  the first  inequality with
$-|d_1|$  (recall  that  $d_3  \ge  0$).   Hence,  replacing
$-|d_1|$ with $|d_1|$ preserves the complete positivity.

Notice that without  loss of generality we  can finally take
$d_2  \ge   d_1$.   This  can   be  shown  as   follows.   A
$\pi/2$-rotation  in  $V$  and $U$  around  the  eigenvector
corresponding to eigenvalue $d_3$ permutes eigenvalues $d_1$
and  $d_2$  (it also  permutes  $c_1$  and $c_2$  and  flips
$c_1$'s sign, but this is irrelevant since $c_1 = c_2 = 0)$.
Hence, without loss of generality we take $d_2 \ge d_1$.

Summarizing,  without  loss  of  generality  for  any  qubit
dihedrally covariant channel we assume  that $D \ge 0$ (that
is, $D$ is  positive semi-definite) with $d_2  \ge d_1$, and
that $c_1 = c_2 = 0$ and $c_3 \ge 0$.

In the setup we consider,  channels that differ by input and
output  unitary  and  anti-unitary  transformations  are  of
course  indistinguishable   in  a   data-driven  way.
Hence,  for any  given  qubit  dihedrally covariant  channel
$\ch{C}_{A, \vec{b}}$,  we will  consider the  qubit channel
$\ch{C}_{D,  \vec{c}}$, with  $D  =  \diag(d_1, d_2,  d_3)$,
where $d_k$'s are the singular values of $A$, and $\vec{c} =
(0, 0, c_3)$, where $c_3 = |\vec{b}|_2$.

\subsection{Binary conditional probability distributions compatible with qubit dihedrally-covariant channel}
\label{sec:compatibility}

Let us now derive  the set $\corr{S}(\ch{C}_{D,\vec{c}})$ of
binary  conditional probability  distributions [that  is, of
  points  $(x,  y)$,  according to  the  parametrization  in
  Eq.~\eqref{eq:cartesian}] that are  compatible with any
given        qubit       dihedrally-covariant        channel
$\ch{C}_{D,\vec{c}}$.  As an  immediate consequence of Lemma
1  of   Ref.~\cite{DBB17},  the   extremal  points   $p$  of
$\corr{S}(\ch{C}_{D,\vec{c}})$  all  satisfy  the  following
condition:
\begin{align}
  \label{eq:compatibility}
  \max_\omega     \left[    p^T     \cdot    w(\omega)     -
    \thr{\omega}{\ch{C}_{D,\vec{c}}} \right] = 0,
\end{align}
where  $w(\omega)$   and  $\thr{\omega}{\ch{C}_{D,\vec{c}}}$
represent  a witness  and  its  threshold, respectively,  and
$\omega$ is a (in general, multidimensional) parameter.

The witness  threshold $\thr{\omega}{\ch{C}_{D,\vec{c}}}$ is
defined as
\begin{align}
  \label{eq:threshold}
  \thr{\omega}{\ch{C}_{D,\vec{c}}}  := \max_{\{\rho_i\},  \{\pi_j\}}
  \sum_{i,           j}           w(\omega)_{i,           j}
  \Tr[\ch{C}_{D,\vec{c}}(\rho_i) \pi_j],
\end{align}
where  the maximization  is  over any  quantum encoding  $\{
\rho_i \}$ and decoding $\{ \pi_j\}$.

For binary conditional probability
distribution   $p$,  as   a  consequence   of  Lemma   2  of
Ref.~\cite{DBB17}, it suffices  to consider diagonal witness
$w(\omega)$, that is
\begin{align*}
  w(\omega) :=  \begin{pmatrix} \frac{1+\omega}2 & 0  \\ 0 &
    \frac{1-\omega}2 \end{pmatrix},
\end{align*}
with $\omega \ge  0$. The cases of  anti-diagonal witness or
$\omega < 0$ also considered in Lemma 2 of Ref.~\cite{DBB17}
can be disregarded without loss  of generality.  This can be
shown as  follows.  Notice first that  the witness threshold
$\thr{\omega}{\ch{C}_{D,\vec{c}}}$                        in
Eq.~\eqref{eq:compatibility} is independent of the choice of
witness  (diagonal  or anti-diagonal)  and  on  the sign  of
$\omega$. Indeed, such choices correspond to permutations of
the  rows   or  columns   of  $w(\omega)$,  which   in  turn
corresponds  to  a relabeling  of  the  optimal encoding  or
decoding.  Moreover,  for a  diagonal witness the  term $p^T
\cdot w(\omega)$ in Eq.~\eqref{eq:compatibility} becomes
\begin{align*}
  p^T  \cdot w(\omega)  =  \frac12  \left( 1  +  y +  \omega
  x\right).
\end{align*}
which, for $\omega > 0$, is maximized by non-negative $x$ or
$y$, respectively, to which  we are restricting without loss
of  generality.  By  explicit computation,  an anti-diagonal
witness or  a negative  $\omega$ lead to  a term  $p^T \cdot
w(\omega)$ which  is maximized by  negative $x$ or  $y$, and
can therefore be disregarded.

It was shown  in Lemma 3 of  Ref.~\cite{DBB17} that the
optimal   encoding  is   orthonormal   (even  for   non
commutativity-preserving  channels),  hence the  witness
threshold is given by
\begin{align*}
  \thr{\omega}{\ch{C}_{D,\vec{c}}} = \max_{\substack{\vec{v}
      \\  |\vec{v}|_2   \le  1}}  \frac12  \left[   1  +  ||
    \ch{C}_{D,\vec{c}}    (\hel{\omega}{\vec{v}}   )    ||_1
    \right],
\end{align*}
where    $\hel{\omega}{\vec{v}}$   denotes    the   Helstrom
matrix~\cite{Hel76} and for qubit channels one has
\begin{align*}
  \ch{C}_{D,\vec{c}}(\hel{\omega}{\vec{v}})   =  \frac{1}{2}
  \left[ \omega \openone + \left( D \vec{v} + \omega \vec{c}
    \right)^T \!\cdot\vec{\sigma} \right],
\end{align*}
whose eigenvalues  are $\left( \omega \pm  \left|D \vec{v} +
\omega  \vec{c}\right|_2  \right)/2$.    Thus,  the  witness
threshold $\thr{\omega}{\ch{C}_{D,\vec{c}}}$  can be readily
computed as
\begin{align*}
  \thr{\omega}{\ch{C}_{D,\vec{c}}} = \frac{1}{2}  \left[ 1 +
    \dst{\omega}{\ch{C}_{D,\vec{c}}} \right].
\end{align*}
where
\begin{align*}
  \dst{\omega}{\ch{C}_{D,\vec{c}}}                        :=
  \max_{\substack{\vec{v}  \\ \left|  \vec{v} \right|_2  \le
      1}} \left|D \vec{v} + \omega \vec{c} \right|_2,
\end{align*}
whenever $\ch{C}_{D,\vec{c}}(\hel{\omega}{\vec{v}})$  is not
semi-definite.                                          When
$\ch{C}_{D,\vec{c}}(\hel{\omega}{\vec{v}})$               is
semi-definite,  the threshold  is  attained  by the  trivial
decoding, hence disregarding this possibility corresponds to
disregarding points  $(x = \pm1, y  = 0)$, that we  will add
back later.

By defining $\vec{u} := D \vec{v}$ one has 
\begin{align*}
  \dst{\omega}{\ch{C}_{D,\vec{c}}}                         =
  \max_{\substack{\vec{u}, \vec{z} \\  \left| D^{-1} \vec{u}
      + \left( \openone -  D^{-1}D \right) \vec{z} \right|_2
      \le 1}} \left| \vec{u} + \omega \vec{c} \right|_2,
\end{align*}
where     $(\cdot)^{-1}$    denotes     the    Moore-Penrose
pseudo-inverse.  Since  vectors $D^{-1} \vec{u}$  and $\left(
\openone  - D^{-1}D  \right)  \vec{z}$  are orthogonal,  the
maximum   is   achieved   by    $\vec{z}   =   0$.    Hence,
$\dst{\omega}{\ch{C}_{D,\vec{c}}}$ is  the maximum Euclidean
distance of vector $-\omega  \vec{c}$ and ellipsoid $|D^{-1}
\vec{u}|_2   \le   1$.

It was shown Lemma 4 of Ref.~\cite{DBB17} that
\begin{align*}
  \dst{\omega}{\ch{C}_{D,\vec{c}}} =
  \begin{cases}
    d_3  +   c_3  \omega,  &   \textrm{  if  }   \omega  \ge
    \omega_0,\\ d_2 \sqrt{1  + \frac{c_3^2 \omega^2}{d_2^2 -
        d_3^2}}, & \textrm{ if } \omega < \omega_0,
  \end{cases}
\end{align*}
where $\omega_0 := (d_2^2  - d_3^2)/(d_3 c_3)$. By replacing
$p^T \cdot w(\omega)$ and $\thr{\omega}{\ch{C}_{D,\vec{c}}}$
into Eq.~\eqref{eq:compatibility} one gets
\begin{align*}
  \max_\omega f(\omega) \le 0, \textrm{ where } f(\omega) :=
  y + \omega x - \dst{\omega}{\ch{C}_{D,\vec{c}}}.
\end{align*}
By explicit computation, the first and second derivatives of
$f$ are given by
\begin{align*}
  \frac{\dif  f}{\dif \omega} (\omega)
  =
  \begin{cases}
    x  - c_3,  & \textrm{  if }  \omega \ge  \omega_0\\ x  -
    \frac{d_2    c_3^2    \omega}{\sqrt{\left(    d_2^2    -
        d_3^2\right)   \left(   c_3^2  \omega^2   +d_2^2   -
        d_3^2\right)}}, & \textrm{ if } \omega < \omega_0.
  \end{cases}
\end{align*}
and
\begin{align*}
  \frac{\dif^2  f}{\dif \omega^2} (\omega)
  =
  \begin{cases}
    0, &  \textrm{ if  } \omega  \ge \omega_0\\  - \frac{d_2
      c_3^2   \sqrt{d_2^2-d_3^2}}{  \left(   c_3^2  \omega^2
      +d_2^2 - d_3^2\right)^\frac32} \le  0, & \textrm{ if }
    \omega < \omega_0.
  \end{cases}
\end{align*}
Hence, by  direct inspection $f(\omega)$ is  continuous with
continuous  first  derivative,  and  concave.   By  explicit
computation, the zero of the first derivative of $f(\omega)$
attained for $\omega \ge 0$ is given by
\begin{align*}
  \omega_1         :=         \frac         {(d_2^2-d_3^2)x}
        {c_3\sqrt{c_3^2d_2^2-(d_2^2-d_3^2)x^2}},
\end{align*}
whenever $\omega_1 < \omega_0$.

By direct  inspection, if $x  \le c_3$ the condition  $0 \le
\omega_1  \le  \omega_0$ is  equivalent  to  $d_2 \ge  d_3$.
Hence,    when    $x    \le    c_3$,    the    maximum    in
Eq.~\eqref{eq:compatibility} is given by
\begin{align*}
  \omega^*  := \begin{cases}  0,  & \textrm{  if  } d_2  \le
    d_3,\\ \omega_1, & \textrm{ if } d_2 > d_3.
    \end{cases}
\end{align*}
By  direct  inspection,  when  $x  >  c_3$  the  maximum  in
Eq.~\eqref{eq:compatibility}  is unbounded,  and hence  this
case can  be disregarded  without loss of  generality. Then,
solving Eq.~\eqref{eq:compatibility}  when $x \le  c_3$, one
has
\begin{align}
  \label{eq:ellipse}
  y  =   \begin{cases}  d_3   &  \textrm{   if  }   d_2  \le
    d_3,\\  \frac1{c_3}\sqrt{d_2^2c_3^2  -  \left(  d_2^2  -
      d_3^2 \right) x^2} & \textrm{ if } d_2 > d_3.
  \end{cases}
\end{align}

By  taking the  symmetric  of Eq.~\eqref{eq:ellipse}  around
axis $x$ and $y$ one recovers the intersection $\corr{E}$ of
an ellipse with the strip $|x| \le c_3$, that is
\begin{align*}
  \corr{E} :=  \left\{ (x,y) \; \Big|  \; (x, y) Q  (x, y)^T
  \le 1  \right\} \cap \left\{ (x,y)  \; \Big| \; |x|  < c_3
  \right\},
\end{align*}
where
\begin{align*}
  Q := \begin{cases} \diag\left( 0, \frac1{d_3^2} \right), &
    \textrm{    if   }    d_2    \le   d_3,\\    \diag\left(
    \frac{d_2^2-d_3^2}{d_2^2c_3^2}, \frac1{d_2^2} \right), &
    \textrm{ if } d_2 > d_3.
  \end{cases}
\end{align*}

Notice  that,  for  any  point $(x,y)$  in  $\corr{E}$,  the
diametrically opposed  point $(-x, -y)$  can be obtained  by a
relabeling of  the decoding, for the  same encoding.  Hence,
any  point  in $\corr{E}$  can  be  obtained without  shared
randomness between the encoding and the decoding.  Moreover,
since  points $(\pm1,  0)$ can  be obtained  by the  trivial
decodings, for any encoding, any point in the convex hull of
$\corr{E}$  with $(\pm1,  0)$ can  also be  obtained without
shared randomness between the encoding and the decoding.

Summarizing,  the  set $\mathcal{S}(\ch{C}_{D,\vec{c}})$  of
binary conditional  probability distributions $p$  [that is,
  points  $(x,y)$,  according   to  the  parametrization  in
  Eq.~\eqref{eq:cartesian}]  that  are compatible  with  any
given      qubit      dihedrally      covariant      channel
$\ch{C}_{D,\vec{c}}$ is  given by the convex  hull of points
$(\pm1, 0)$ with $\corr{E}$, that is
\begin{align*}
  \mathcal{S}(\ch{C}_{D,\vec{c}})  = \conv\left[  (\pm1, 0),
    \; \corr{E} \right].
\end{align*}

Additionally,   as    a   consequence   of   Lemma    4   of
Ref.~\cite{DBB17}, for $|x| < c_3$  one has that the optimal
encoding is  given by  Eq.~\eqref{eq:bloch} with  $\vec{v} =
\vec{v}^*$ given by
\begin{align*}
  \vec{v}^*  = \begin{cases}  \left(  0, 0,  \pm1 \right)  &
    \textrm{          if          }         d_2          \le
    d_3,\\  \frac1{\sqrt{d_2^2(c_3^2-x^2)+d_3^2x^2}}\left( 0,
    d_2 \sqrt{c_3^2 - x^2}, d_3 x\right) & \textrm{ if } d_2
    > d_3,
  \end{cases}
\end{align*}
while, again for $|x| <  c_3$, the optimal decoding is given
by Eq.~\eqref{eq:bloch} with $\vec{v} = \vec{u}^*$ given by
\begin{align*}
  \vec{u}^*  = \begin{cases}  \left(  0, 0,  \pm1 \right)  &
    \textrm{  if  }  d_2   \le  d_3,\\  \frac1{c_3}\left(  0,
    \sqrt{c_3^2 - x^2}, x\right) & \textrm{ if } d_2 > d_3.
  \end{cases}
\end{align*}

\subsection{Equivalence classes of data-drivenly indistinguishable qubit dihedrally-covariant channels}
\label{sec:equivalence}

Here we derive the equivalence  classes of channels that are
data-drivenly  indistinguishable.   By defining  with
$V(\ch{C}_{D,\vec{c}})$         the        volume         of
$\corr{S}(\ch{C}_{D,\vec{c}})$ (for example, with respect to
the flat  Euclidean metric), the  qubit dihedrally-covariant
channel $\ch{C}_{D^*,\vec{c}^*}$ compatible with the minimal
volume  in  the  space  of  binary  conditional  probability
distributions that  is also compatible with  observed binary
conditional probability distributions  $\corr{D} := \{ (x_k,
y_k) \}_k$ is given by
\begin{align}
  \label{eq:opt}
  \left(          D^*,\vec{c}^*           \right)          =
  \arg\min_{\substack{D,\vec{c}\\c_1  =  c_2   =  0\\d1  \le
      d2\\Eq.~\eqref{eq:cp}\\\corr{D}              \subseteq
      \corr{S}(\ch{C}_{D,\vec{c}})}} V(\ch{C}_{D,\vec{c}}).
\end{align}

Once  any   such  a   qubit  dihedrally   covariant  channel
$\ch{C}_{D,\vec{c}}$  has been  found,  it  is important  to
characterize   the    class   of   equivalence    of   qubit
dihedrally-covariant    channels    that    also    minimize
Eq.~\eqref{eq:opt}. Upon defining
\begin{align*}
  \mu  \left(   \ch{C}_{D,\vec{c}}  \right)   :=  \frac{1-c_3}{c_3}
  \frac{d_2^2-d_3^2}{d_3^2},
\end{align*}
by direct inspection of  Eq.~\eqref{eq:ellipse} one has that
$x =  c_3$ (in which  $y = d_3$)  is a discontinuity  of the
derivative         of         the        boundary         of
$\mathcal{S}(\ch{C}_{D,\vec{c}})$    if    and    only    if
$\mu(\ch{C}_{D,\vec{c}})  \le 1$,  in which  case $c_3$  and
$d_3$   are    invariants   of   the    equivalence   class.
Additionally,    again    by   Eq.~\eqref{eq:ellipse},    if
$\mu(\ch{C}_{D,\vec{c}})    >    0$,    also    $d_2$    and
$(d_2^2-d_3^2)/(d_2^2c_3^2)$  are  invariants.  Finally,  if
$c_3  = c_3'  = 0$  or  $d_3 =  d_3'  = 0$  [in which  cases
  $\mu(\mathcal{C}_{D, \vec{c}})$  and $\mu(\mathcal{C}_{D',
    \vec{c}'})$  are  undefined],   the  only  invariant  is
$\max(d_2, d_3)$.

Summarizing,   two   qubit   dihedrally-covariant   channels
$\mathcal{C}_{D, \vec{c}}$  and $\mathcal{C}_{D', \vec{c}'}$
are data-drivenly indistinguishable if and only if:
\begin{description}
\item[Regime
  $\mu(\ch{C}_{D,\vec{c}}),   \mu(\ch{C}_{D',\vec{c}'})
  \le 0$ ]
  \begin{align*}
    \begin{cases}
      d_3 = d_3',\\
      c_3 = c_3'.
    \end{cases}
  \end{align*}
\item[Regime
  $0             <             \mu(\ch{C}_{D,\vec{c}}),
  \mu(\ch{C}_{D',\vec{c}'}) < 1$]
\begin{align*}
  \begin{cases}
    d_2 = d_2',\\
    d_3 = d_3',\\
    c_3 = c_3'
  \end{cases}
\end{align*}
\item[Regime
  $1      \le       \mu(\mathcal{C}_{D,      \vec{c}}),
  \mu(\mathcal{C}_{D', \vec{c}'})$]
  \begin{align*}
  \begin{cases}
    d_2 = d_2',\\
    \frac{d_2^2-d_3^2}{c_3^2} = \frac{d_2'^2-d_3'^2}{c_3'^2},\\
  \end{cases}
\end{align*}
\item[Regime $c_3 = c_3' = 0$ (Pauli channel)]
  \begin{align*}
    \max\left(d_2, d_3\right) =     \max\left(d_2', d_3'\right).
  \end{align*}
\end{description}

Outside    of    the     aforementioned    regimes,    qubit
dihedrally-covariant  channels   $\ch{C}_{D,  \vec{c}}$  and
$\ch{C}_{D',     \vec{c}'}$     are     data-drivenly
distinguishable.

\end{document}